\documentclass[runningheads]{llncs/llncs}

\usepackage{graphicx} % Required for inserting images
\usepackage{amsmath}
\usepackage{amssymb}
\usepackage{extarrows}
\usepackage[ruled,linesnumbered]{algorithm2e}
\usepackage{cleveref}
\usepackage{tikz}
\usepackage{pgffor}
\usepackage{multirow}
\usepackage{mathpartir}
\usepackage{subcaption}
\usetikzlibrary{arrows.meta, fit, automata, positioning, calc, trees, shapes.multipart}

\SetKwComment{Comment}{/* }{ */}

\newcommand{\sqin}{\sqsubset\!\!\!\!\!\!\!{-}~}

\newcommand{\true}{\textit{true}}
\newcommand{\false}{\textit{false}}

\newcommand{\bisim}{\,\underline{\leftrightarrow}\,}

\newcommand{\nomono}{A^*_{\setminus m}}

\iftrue
\newcommand{\emptydiamond}{%
\begin{tikzpicture}%
\draw[line cap=round, line join=round] (-.1ex,-.1ex) -- (1.1ex,1.1ex);%
\draw[line cap=round, line join=round] (0ex,.5ex) -- (.5ex,0ex) -- (1ex, .5ex) -- (.5ex, 1ex) -- (0ex,.5ex);
\end{tikzpicture}%
}
\fi

\newcounter{subexample}[example]
\renewcommand{\thesubexample}{(\arabic{example}\alph{subexample})}
\newcommand{\subexamplecaption}[1]{
    \newline
    \refstepcounter{subexample}
    \thesubexample #1
}

\title{Mining Diamonds in Labelled Transition Systems}
\author{P.H.M.\ van Spaendonck \orcidID{0000-0002-9536-1524}    \thanks{This publication is part of the PVSR project (with project number 17933) of the MasCot research programme which is financed by the Dutch Research Council (NWO).} \\
\and  K.H.J.\ Jilissen \orcidID{0000-0002-4697-2011}
}
\institute{
    Eindhoven University of Technology \\
    \email{p.h.m.v.spaendonck@tue.nl, k.h.j.jilissen@tue.nl}
}
\date{}

\begin{document}

\maketitle
\begin{abstract}
    Labelled transition systems can be a great way to visualise the complex behaviour of parallel and communicating systems.
    However, if, during a particular time frame, no synchronisation or communication between processes occurs, then multiple parallel sequences of actions are able to interleave arbitrarily, and the resulting graph quickly becomes too complex for the human eye to understand easily.

    With that in mind, we propose an exact formalisation of these arbitrary interleavings, and an algorithm to find all said interleavings in any arbitrary finite LTS, to reduce the visual complexity of labelled transition systems.
\end{abstract}
\keywords{Labelled Transition Systems $\cdot$ State Space Reduction $\cdot$ Concurrent Systems $\cdot$ State Space Visualisation}

\section{Introduction}
Parallel and communicating systems are often difficult to understand due to the divergence that arises from asynchronicity.
Toward this end, labelled transition systems (LTSs) are a common way to visualise the divergent behaviour of such systems.
As with many visualisation techniques, keeping the visual complexity of a given LTS at a minimum is important for its effectiveness as a communication tool between model-based engineers and software engineers.
Techniques for reducing this visual complexity are not a novel field and have been studied extensively, e.g.\ reduction through equivalence relations or pre-orders.

We identify a pattern that can occur easily in communication systems and which introduces large amounts of visual clutter.
That is, if, during a particular time frame, no synchronisation or communication between processes occurs, then multiple parallel sequences of actions can interleave arbitrarily, and the resulting graph quickly becomes too complex for the human eye to understand easily.
In particular, we argue that while the fact that no synchronisation or communication occurs is useful information for an engineer, given that the result of each interleaving is always the same, explicitly labelling each possible interleaving is merely an exercise in combinatorics.

This pattern, dubbed a \emph{diamond pattern} due to its structure, occurs when we have a set of asynchronous sequential actions that must all fully execute before any further progress can occur.
For example, work is divided up over multiple threads and calculations can only continue after all threads have calculated their results.
Given $n$ parallel processes, the pattern itself becomes a $n$-dimensional hypercube that consists of all possible interleavings of these sequences, cf.\ the two LTSs in Example \ref{fig:simple-diamond-pattern-l} and \ref{fig:simple-diamond-pattern-m}.
The aforementioned example is still relatively easy to understand, but with the number of parallel processes increasing, this is no longer the case.

\begin{example} 
    \label{fig:simple-diamond-pattern}
    The two LTSs \ref{fig:simple-diamond-pattern-l} and \ref{fig:simple-diamond-pattern-m} both contain all possible interleavings of the sequences $a_1\cdot a_2$ and $b$, starting from state $\hat{q}$. 
    The LTS \ref{fig:simple-diamond-pattern-r} contains a single ``macro''-transition denoting that the sequence $a_1 \cdot a_2$ occurs in parallel to the sequence $b$.

    \centering
    \begin{minipage}{0.3\textwidth}
    \centering
    \begin{tikzpicture}
    \node (q00) at (0,0) {$\hat{q}$};
    \node (q01) at (.75,-.75) {$.$};
    \node (q02) at (1.5,-1.5) {$.$};
    \node (q10) at (-.75,-.75) {$.$};
    \node (q11) at (0,-1.5) {$.$};
    \node (q12) at (.75,-2.25) {$\check{q}$};

    \draw[->] (q00) -- node[pos=.5,  above] {$a_1$} (q01);
    \draw[->] (q01) -- node[pos=.5,  above] {$a_2$} (q02);
    \draw[->] (q00) -- node[pos=.5,  above] {$b$} (q10);
    \draw[->] (q01) -- node[pos=.5,  above] {$b$} (q11);
    \draw[->] (q02) -- node[pos=.5,  above] {$b$} (q12);
    \draw[->] (q10) -- node[pos=.5,  above] {$a_1$} (q11);
    \draw[->] (q11) -- node[pos=.5,  above] {$a_2$} (q12);
    
    \end{tikzpicture}
    \subexamplecaption{}
    \label{fig:simple-diamond-pattern-l}
    \end{minipage}
    \begin{minipage}{0.3\textwidth}
    \centering
    \begin{tikzpicture}
    \node (qe) at (0,0) {$\hat{q}$};
    
    \node (qb) at (-.75, -.75) {$.$};
    \node (qa1) at (.75,-.75) {$.$};
    
    \node (qba1) at (-.75, -1.5) {$.$};
    \node (qa1b) at (0, -1.5) {$.$};
    \node (qa1a2) at (.75,-1.5) {$.$};

    \node (qf) at (0, -2.25) {$\check{q}$};

    \draw[->] (qe) -- node[pos=.5, above] {$b$} (qb);
    \draw[->] (qe) -- node[pos=.5, above] {$a_1$} (qa1);
    \draw[->] (qb) -- node[pos=.5, left] {$a_1$} (qba1);
    \draw[->] (qa1) -- node[pos=.5, above] {$b$} (qa1b);
    \draw[->] (qa1) -- node[pos=.5, right] {$a_2$} (qa1a2);
    \draw[->] (qba1) -- node[pos=.5, left] {$a_2$} (qf);
    \draw[->] (qa1b) -- node[pos=.2, left] {$a_2$} (qf);
    \draw[->] (qa1a2) -- node[pos=.5, above] {$b$} (qf);
    \end{tikzpicture}
    \subexamplecaption{}
    \label{fig:simple-diamond-pattern-m}
    \end{minipage}
    \begin{minipage}{0.15\textwidth}
    \centering
    \begin{tikzpicture}
    \node (qhat) at (0,0) {$\hat{q}$};
    \node (qcheck) at (0,-2.25) {$\check{q}$};

    \draw[->] (qhat) -- node[pos=.5, right] {$a_1a_2 || b$} (qcheck);
    \end{tikzpicture}
    \subexamplecaption{}
    \label{fig:simple-diamond-pattern-r}
    \end{minipage}
\end{example}

The difficulty in noticing these patterns is not only limited to cases with a large number of parallel processes.
In Example \ref{ex:strange-diamonds}, the two LTSs have near identical structures.
The LTS \ref{fig:strange-diamonds-l} can be summarised as the interleaving $ba||ca$, but the LTS \ref{fig:strange-diamonds-r} cannot.
Note also that all interleavings end in the state $\check{q}$, yet the grid-like structure often associated with such congruences ends in the state right before it.

\begin{example} \label{ex:strange-diamonds}
    The two LTSs below have near identical structures, the only difference being the label of the final transitions to $\check{q}$ and $\check{p}$.
    However, only the LTS \ref{fig:strange-diamonds-l} contains all (and only all) interleaving of the two sequences $b \cdot a$ and $c \cdot a$, while in the LTS \ref{fig:strange-diamonds-r} a second action $a$ is never enabled.

    \centering
    \begin{minipage}{0.25\textwidth}
    \centering
    \begin{tikzpicture}
    \node (q1) at (0,0) {$\hat{q}$};
    \node (q2) at (-.5, -.75) {.};
    \node (q3) at (.5, -.75) {.};
    \node (q4) at (-1, -1.5) {.};
    \node (q5) at (0, -1.5) {.};
    \node (q6) at (1, -1.5) {.};
    \node (q7) at (0, -2.25) {.};
    \node (q8) at (0, -3) {$\check{q}$};

    \draw[->] (q1) -- node[pos=.2,left] {$b$} (q2);
    \draw[->] (q1) -- node[pos=.2,right] {$c$} (q3);
    \draw[->] (q2) -- node[pos=.2,left] {$a$} (q4);
    \draw[->] (q2) -- node[left] {$c$} (q5);
    \draw[->] (q3) -- node[right] {$b$} (q5);
    \draw[->] (q3) -- node[pos=.2,right] {$a$} (q6);
    \draw[->] (q4) -- node[left] {$c$} (q7);
    \draw[->] (q5) -- node[pos=.2,left] {$a$} (q7);
    \draw[->] (q6) -- node[right] {$b$} (q7);
    \draw[->] (q7) -- node[left] {$a$} (q8);
    \end{tikzpicture}
    \subexamplecaption{}
    \label{fig:strange-diamonds-l}
    \end{minipage}
    \begin{minipage}{0.25\textwidth}
    \centering
    \begin{tikzpicture}
    \node (q1) at (0,0) {$\hat{p}$};
    \node (q2) at (-.5, -.75) {.};
    \node (q3) at (.5, -.75) {.};
    \node (q4) at (-1, -1.5) {.};
    \node (q5) at (0, -1.5) {.};
    \node (q6) at (1, -1.5) {.};
    \node (q7) at (0, -2.25) {.};
    \node (q8) at (0, -3) {$\check{p}$};

    \draw[->] (q1) -- node[pos=.2,left] {$b$} (q2);
    \draw[->] (q1) -- node[pos=.2,right] {$c$} (q3);
    \draw[->] (q2) -- node[pos=.2,left] {$a$} (q4);
    \draw[->] (q2) -- node[left] {$c$} (q5);
    \draw[->] (q3) -- node[right] {$b$} (q5);
    \draw[->] (q3) -- node[pos=.2,right] {$a$} (q6);
    \draw[->] (q4) -- node[left] {$c$} (q7);
    \draw[->] (q5) -- node[pos=.2,left] {$a$} (q7);
    \draw[->] (q6) -- node[right] {$b$} (q7);
    \draw[->] (q7) -- node[left] {$d$} (q8);
    \end{tikzpicture}
    \subexamplecaption{}
    \label{fig:strange-diamonds-r}
    \end{minipage}
\end{example}

As such, this paper aims to find these \emph{diamond patterns} in arbitrary (non-deterministic) LTSs without any prior knowledge about the relation between its action labels, and replacing them with a single transition that captures the precise parallelisation that occurs, e.g.\ see the LTS in Example \ref{fig:simple-diamond-pattern-r}.
Especially towards reducing the visual complexity of arbitrary LTSs, finding the specific parallelised sequences can be useful to highlight aspects of the modelled system.

Specifically, the contributions of this paper are as follows:
We present a formalisation of the diamond pattern, including proofs on the correctness of the formalisation. 
We investigate the relationship between diamond convergences and equivalence relations on LTSs and prove that strong bisimilarity is the exact relation that preserves diamond convergences for arbitrary LTSs.
Last, we present an algorithm for finding all diamonds in any arbitrary LTS.

In Section \ref{sec:related-work}, we briefly discuss the closely related body on partial order refinement and other techniques aimed at simplifying the visualisation of complex communicating systems.
In Section \ref{sec:preliminary}, we give the definitions of and related to LTSs that are used throughout the rest of the paper.
In Section \ref{sec:definitions} we formalise what diamonds are and when a diamond occurs in an LTS and prove some rudimentary properties aimed at the completeness and soundness of our definitions.
In Section \ref{sec:diamond-preservation} we prove that strong bisimilarity is the precise equivalence relation that preserves diamond convergence.
%Thus, we can safely reduce an arbitrary LTS modulo strong bisimilarity without removing or adding any diamonds.
In Section \ref{sec:finding-diamonds} we outline our novel algorithm for finding all diamonds in any LTS and outline its correctness proof.
In Section \ref{sec:conclusion} we conclude and discuss some possible directions for future work related to diamond patterns.

\section{Related Work} \label{sec:related-work}
The work presented here is closely related to partial order reduction (POR) methods, e.g.\ \cite{baier2008principles,godefroid1996partial,10.1007/3-540-56922-7_34,valmari1991stubborn}, in which the commutativity of parallel and independent operations is used to bundle together functionally equivalent traces.
These POR methods are used during state space exploration and model checking to avoid having to consider multiple functionally equivalent traces, and thus accelerate said algorithms.
To achieve this speedup, POR methods have to consider only a strict subset of all traces/transitions.
Which subset is considered influences the set of properties that is preserved, e.g.\ LTL for ample sets \cite{10.1007/3-540-56922-7_34}, ${\text{CTL}^*}_{\setminus \circ}$ for ample sets with stuttering \cite{baier2008principles}, or deadlock for stubborn sets \cite{valmari1991stubborn}.
As we show later, reduction modulo diamond convergence is equivalent to reduction modulo strong bisimulation. It is thus much stronger in preserving behaviour than the aforementioned POR techniques by preserving all properties in $\text{CTL}^*$\cite{grumberg1999model}, and only matched when strong assumptions are made a-priori about the parallelisation of the system, e.g.\ in \cite{huhn1998partial}.

As noted by Groote et al.\ in \cite{groote2016random}, these convergences are inherent to any parallel systems.
However, POR techniques have to consider only a strict subset of these convergences to gain any potential speedup during state space exploration.
More-so, they can only consider outgoing transitions to find these, whereas the reduction algorithm introduced here finds all diamond convergences by traversing incoming transitions.

We now discuss some other methods for reducing the visual complexity of LTSs, since our work is aimed towards doing the same.
In \cite{van2021visualisation}, van Ieperen presents layout and rendering techniques aimed at improving the visual readability of LTSs with over 10000 states.
These techniques are applied to various large pre-existing formal models.
Of particular interest are the presented Alma, leader election protocol \cite{10.1145/357195.357200}, and Twilight models, which contain large degrees of asynchronous behaviour, and thus contain many diamond patterns and/or patterns similar to diamond patterns.
However, as the authors themselves note, a survey with a statistically significant sample size is still required to properly assess the effectiveness of these techniques.

In \cite{herman2000graph}, Herman et al.\ conduct a significantly large survey on various visualisation techniques for graphs.
These visualisation techniques are specific to visualising parts of a graph as a tree with a single begin- and multiple endpoints.
However, we note that communicating systems can often be cyclic, such as the aforementioned Alma and Twilight models in \cite{van2021visualisation}, and thus these techniques can only be applied to specific, non-cyclic, parts of particular models.

In \cite{groote2006interactive}, Groote and van Ham present the LTSView tool.
This tool, aimed at providing insight into the global structure of significantly large statespaces, clusters together similar states into discs and arranges these discs into a 3-dimensional treelike structure.
The authors use the tool to provide information on various real world models, including a model with just over $10^6$ states.

The visual complexity of an LTS can also be improved by reducing the size of a given LTS.
This can be done by reducing modulo some equivalence relation, e.q.\ trace equivalence, strong bisimulation \cite{10.1007/BFb0017309}, or branching bisimulation \cite{van1996branching}, or modulo some preorder relation, e.g.\ the simulation preorder \cite{cleaveland2001equivalence}.

It should also be noted that LTSs are not the only way to visualise complex parallel systems.
Other visualisation/formalisation techniques include, in no particular order, Kripke Structures \cite{kripke1963semantical}, Petri Nets \cite{peterson1977petri}, and state transition graphs \cite{pretorius2006visual,van2001visualization}, and variations of all these.
Each technique focuses on different aspects of communicating behaviour, and thus each has its own respective advantages and disadvantages.

\section{Preliminaries} \label{sec:preliminary}
A \emph{labelled Transition System} is a simple way to formalise the behaviour of a non-deterministic (parallel) system as a directed graph \cite{milner1999communicating}.
Edges are labelled with actions taken from some action-alphabet $A$, representing some atomic event occuring, and lead to a (possibly new) state.
For this paper, we only consider LTSs with a finite set of states and action labels.

\begin{definition} \label{def:lts}
    A labelled Transition System (LTS) is defined as a tuple \\ $\langle Q, q_0, A, \to \rangle$ where:
    \begin{itemize}
        \item $Q$ is the finite set of states;
        \item $q_0 \in Q$ is the initial state;
        \item $A$ is a finite set of action labels;
        \item $\to \subseteq Q \times A \times Q$ is the transition relation where $\langle q, a, q' \rangle \in \to$ is usually written as $q \xrightarrow{a} q'$.
    \end{itemize}
    Additionally, given any LTS $\langle Q, q_0, A, \to \rangle$, state $q \in Q$, and label $a \in A$, we have $q \xrightarrow{a} \overset{def}{=} \exists_{q'\in Q}[q \xrightarrow{a} q']$.
\end{definition}

Throughout the paper, we make use of \emph{action sequences}, sometimes referred to as words or strings in automata theory \cite{hopcroft2001introduction,10.5555/1096945}, which are concatenations of zero or more actions.

\begin{definition} \label{def:action-sequences}
    Given a set of actions $A$, the set $A^*$ denotes the set of action sequences.
    We use $\varepsilon \in A^*$ to denote the empty sequence.
    Additionally, given any action $a \in A$ and action sequences $s \in A^*$, we have: 
    \begin{itemize}
        \item $\textit{hd}(as) = a$, $\textit{tl}(as) = s$,
        \item the length $|~| : A^* \to \mathbb{N}$ is defined s.t.\ $|\varepsilon| = 0$ and $|as| = 1 + |s|$, and
        \item the minimal alphabet $A^- : A^* \to \mathcal{P}(A)$ is defined s.t.\ $A^-(\varepsilon) = \emptyset$ and $A^-(as) = \{a\} \cup A^-(s)$.
    \end{itemize}
    Additionally, we generalise the $\to$ relation for LTSs s.t.\ given any LTS 
    $\langle Q, q_0, A,$ $ \to \rangle$, we have $q \xrightarrow{\varepsilon} q$ for all $q \in Q$, we have $q \xrightarrow{as} q''$ for all $q,q'' \in Q, a \in A, s \in A^*$ iff there is some state $q' \in Q$ with $q \xrightarrow{a} q' \xrightarrow{s} q''$, and we have $q \xrightarrow{s}$ for all $q \in Q, s\in A^*$ iff there is some state $q' \in Q$ with $q \xrightarrow{s} q'$.
\end{definition}

As noted earlier, notions of equivalences of different LTSs have been studied extensively, e.g.\ the work by Van Glabbeek \cite{van2001linear} on the linear time-branching time spectrum gives a nice overview on various equivalence relations.
Two common equivalence relations, of which we discuss their relation to the diamond patterns later, are \emph{trace equivalence} \cite{olderog1986specification}, and \emph{strong bisimilarity} \cite{milner1992functions,park2005concurrency}.

\begin{definition} \label{def:trace-equiv}
      Given two LTSs $l = \langle Q, q_0, A, \to \rangle$, and $l' = \langle Q', q_0', A, \to' \rangle$, we say that $l$ and $l'$ are trace equivalent, denoted as $l =_{tr} l'$, iff $\textit{traces}(q_0) = \textit{traces}(q_0')$, where for any state $q \in Q$, $\textit{traces}(q) = \{ s \in A^* | q \xrightarrow{s}\}$.
\end{definition}

\begin{definition} \label{def:bisim}
    Given an LTS $\langle Q, q_0, A, \to \rangle$ and some relation $R \subseteq Q \times Q$, we say that $R$ is a bisimulation iff:
    \begin{itemize}
        \item $R$ is symmetric;
        \item Given states $q,q',q_{to}\in Q$ and action $a \in A$ such that $q \xrightarrow{a} q_{to}$ and $q{R}q'$ then there is some state $q_{to}' \in Q$ with $q' \xrightarrow{a} q_{to}'$ and $q_{to}{R}q_{to}'$.
    \end{itemize}
    We say two states $q,q' \in Q$ are (strongly) bisimilar, denoted as $q \bisim q'$, iff there is a bisimulation $R$ with $q{R}q'$.\\
    We say two LTSs $l = \langle Q, q_0, A, \to \rangle$ and $l' = \langle Q', q_0, A, \to' \rangle $ are (strongly) bisimilar iff there is bisimulation $R$ s.t.\ $q_0{R}q_0'$ in $\langle Q \uplus Q', q_0, A \cup A', \to \cup \to' \rangle$.
\end{definition}

Particularly useful when reducing the complexity of an LTS is the notion of some LTS that is minimal modulo a given equivalence relation to describe an LTS in a given equivalence class of LTSs, i.e.\ all LTSs that are equivalent modulo said equivalence relation, that cannot be reduced to have fewer states.
Note that it is not important to us that the number of transitions cannot be further reduced.

\begin{definition} \label{def:minimal-lts}
    Given any equivalence relation $=_x$ on LTSs, and an LTS $l = \langle Q, q_0, A, \to \rangle$.
    We say that $l$ is minimal modulo $=_x$ iff given states $q,q' \in Q$ if $q =_x q'$ then $q = q'$.
\end{definition}

\begin{example}
    \label{ex:simple-equivalence}
    The following three LTSs, with initial states $p, q$, and $r$, are all trace equivalent.
    However the states $q$ and $r$ are not strongly bisimilar, since after the taking the left $a$ transition in $q$, the $c$ action, which is always enabled after taking the $a$ transition in $r$, cannot be simulated.
    The LTS with intial state $q$ is minimal modulo strong bisimulation, and the LTS with initial state $r$ is minimal modulo trace equivalence.

    \centering
    \begin{minipage}{0.3\textwidth}
    \centering
    \begin{tikzpicture}
        \node (q1) at (0,0) {$p$};
        \node (q2) at (-.5, -.75) {.};
        \node (q3) at (-.5, -1.5) {.};
        \node (q4) at (.5, -.75) {.};
        \node (q5) at (.5, -1.5) {.};
        \draw[->] (q1) -- node[pos=.2, left] {$a$} (q2);
        \draw[->] (q2) -- node[left] {$b$} (q3);
        \draw[->] (q3) -- ++(-.5,0) -- node[left] {$c$} ++(0,1.5) -- (q1);
        \draw[->] (q1) -- node[pos=.2,right] {$a$} (q4);
        \draw[->] (q4) -- ++(.25,0) -- node[right] {$c$} ++ (0,.75) -- (q1);
        \draw[->] (q4) -- node[right] {$b$} (q5);
        \draw[->] (q5) -- ++(-.5,0) -- node[left] {$c$} (q1);
    \end{tikzpicture}
    \subexamplecaption{}
    \label{fig:simple-equivalence-l}
    \end{minipage}
    \begin{minipage}{0.3\textwidth}
    \centering
    \begin{tikzpicture}
        \node (p1) at (4,0) {$q$};
        \node (p2) at (3.5, -.75) {.};
        \node (p3) at (4.5, -.75) {.};
        \node (p4) at (4, -1.5) {.};
        \draw[->] (p1) -- node[pos=.2, left] {$a$} (p2);
        \draw[->] (p1) -- node[pos=.2, right] {$a$} (p3);
        \draw[->] (p3) -- ++(.25, 0) 
            -- node[right] {$c$} ++(0,.75) 
            --  (p1);
        \draw[->] (p2) -- node[pos=.4, left] {$b$} (p4);
        \draw[->] (p3) -- node[pos=.4, right] {$b$} (p4);
        \draw[->] (p4) -- node[left] {$c$} (p1);
    \end{tikzpicture}
    \subexamplecaption{}
    \label{fig:simple-equivalence-m}
    \end{minipage}
    \begin{minipage}{0.3\textwidth}
    \centering
    \begin{tikzpicture}
        \node (r1) at (7,0) {$r$};
        \node (r2) at (7, -.75) {.};
        \node (r3) at (7, -1.5) {.};
        \draw[->] (r1) -- node[right] {$a$} (r2);
        \draw[->] (r2) -- ++(.5,0) 
            -- node[right] {$c$} ++(0,.75)
            -- (r1);
        \draw[->] (r2) -- node[right] {$b$} (r3);
        \draw[->] (r3) -- ++(-.35,0) 
            -- node[left] {$c$} ++ (0,1.5) 
            -- (r1);
    \end{tikzpicture}
    \subexamplecaption{}
    \label{fig:simple-equivalence-r}
    \end{minipage}
\end{example}

\section{Diamond Patterns} \label{sec:definitions}
Before we proceed to define what a diamond pattern is, we note a particular oddity, shown in Example \ref{ex:monotone-diamond}, that arises with monotone sequences, i.e.\ a sequence consisting only of the repetition of a single action, e.g.\ $a$, $bbb$.
In such instances, the particular parallel composition leading to the diamond pattern in question, has to be arbitrated post reduction modulo strong bisimulation.
Since reduction modulo some equivalence relation weaker than isomorphism is a commonly applied technique when working with LTSs, we opt to group such monotone (sub-)patterns together in our definition.

\begin{example} \label{ex:monotone-diamond}
The two LTSs below, \ref{fig:monotone-diamond-l} and \ref{fig:monotone-diamond-r}, are strongly bisimilar.
Starting in their initial states $\hat{q}$ and $\hat{p}$, it is possible to perform any interleaving of the sequences $aa$ and $a$.
However, it is also possible to perform any interleaving of the three identical sequences $a$, or simply perform the sequence $aaa$.
After reduction modulo strong bisimulation, the parallel characterisation is ambiguous.

\centering
\begin{minipage}{0.3\textwidth}
\centering
\begin{tikzpicture}
    \node (q1) at (0,0) {$\hat{q}$};
    \node (q2) at (-1.5, -.6) {.};
    \node (q3) at (.6, -1.5) {.};
    \node (q4) at (-.9,-2.1) {.};

    \draw[->] (q1) -- node[above] {$a$} (q2);
    \draw[->] (q1) -- node[right] {$a$} (q3);
    \draw[->] (q2) -- node[pos=.15,right] {$a$} (q4);
    \draw[->] (q3) --node[above] {$a$} (q4);
    
    \node (q5) at (0, -1) {.};
    \node (q6) at (-1.5,-1.6) {.};
    \node (q7) at (.6, -2.5) {.};
    \node (q8) at (-.9, -3.1) {$\check{q}$};

    \draw[->] (q5) -- node[above] {$a$} (q6);
    \draw[->] (q5) -- node[pos=.15,right] {$a$} (q7);
    \draw[->] (q6) -- node[left] {$a$} (q8);
    \draw[->] (q7) --node[above] {$a$} (q8);

    \draw[->] (q1) -- node[left] {$a$} (q5);
    \draw[->] (q2) -- node[left] {$a$} (q6);
    \draw[->] (q3) -- node[right] {$a$} (q7);
    \draw[->] (q4) -- node[right] {$a$} (q8);
\end{tikzpicture}
\subexamplecaption{}
\label{fig:monotone-diamond-l}
\end{minipage}
\begin{minipage}{0.3\textwidth}
\centering
\begin{tikzpicture}
    \node (p0) at (0,0) {$\hat{p}$};
    \node (p1) at (0, -1) {.};
    \node (p2) at (0, -2) {.};
    \node (p3) at (0, -3.1) {$\check{p}$};
    \draw[->] (p0) -- node[right] {$a$} (p1);
    \draw[->] (p1) -- node[right] {$a$} (p2);
    \draw[->] (p2) -- node[right] {$a$} (p3);
\end{tikzpicture}
\subexamplecaption{}
\label{fig:monotone-diamond-r}
\end{minipage}

\end{example}

In Definition \ref{def:diamond}, we define what a \emph{diamond pattern} is.
Simply put, a diamond consists of interleaving actions, representing monotone sequences, and non-monotone action sequences of arbitrary positive lengths.

\begin{definition} \label{def:diamond}
A diamond pattern $\diamond = \langle C_{a}, C_{s} \rangle$ is a tuple of two mappings: $C_{a} \colon A \rightarrow \mathbb{N}$ denoting the cardinality of actions in the diamond, and $C_{s} \colon A^*_{\backslash m} \rightarrow \mathbb{N}$ denoting the cardinality of non-monotone sequences in the diamond, where $A^*_{\backslash m} \subset A^*$ is defined as $A^*_{\backslash m} = \lbrace s \in A^* \mid |A^-(s)| \geq 2 \rbrace$.
The set of all diamond patterns over $A$ is $\Diamond(A)$.

\noindent Additionally, we have:
    \begin{itemize}
        \item we denote the empty diamond, i.e.\ the diamond with cardinality $0$ for all actions and actions sequences, as $\emptydiamond$,
        \item $\textit{hd}(\langle C_a, C_s \rangle) = \{a \in A ~|~ C_a(a) \neq 0 \} \cup \{a \in A ~|~ \exists s \in A^*_{\setminus m}. \textit{hd}(s) = a \wedge C_s(s) \neq 0 \}$, and
        \item the size of a diamond $|~| : \Diamond(A) \to \mathbb{N}$ is defined s.t.\\ $|\langle C_a, C_s \rangle | = \sum_{a \in A} C_a(a) + \sum_{s \in A^*_{\setminus m}} |s|*C_s(s)$.
    \end{itemize}
    
\end{definition}

We often forgo writing down the mappings and instead write down the explicit actions and action sequences that are not mapped to $0$, e.g.\ given $A = \{a,b,c\}$ we write the diamond $\langle C_a, C_s \rangle$ where $C_a = \{a \mapsto 3, b \mapsto 0, c \mapsto 1\}$, and $C_s(ab) = 2, C_s(bcc) = 1$, and $C_s(s) = 0$ for any $s \notin \{ab, bcc \}$ as:
$$ a^3 || c^1 || (ab)^2 || (bcc)^1$$
The diamond above describes the interleaving of $3$ $a$ actions, $1$ $c$ action, $2$ $ab$ sequences, and $1$ $bcc$ sequence.
We later define when such a diamond pattern actually occurs in an LTS.

The \emph{tail} operations, see Definition \ref{def:diamond-tl}, give all possible resulting diamonds after taking a single action, an action sequence, or a diamond.
If the action, action sequence, or diamond is not contained in the diamond, then an empty set is returned.
The \emph{tail} operation for removing a single action is best understood as distinguishing 3 cases: the action $a$ is present one or more times as a single action, the diamond contains one or more sequences $as$ where $s$ is not monotone, and/or the diamond contains one more sequences $as$, where $s$ is monotone.
%We note that, for deterministic diamonds, i.e.\ diamonds of which the action sequences do not share actions, the resulting sets of these mappings always consist of a single diamond given that the actions are enabled.
We show the results of different tail operations on various diamonds in Example \ref{ex:diamond-tl}.

\begin{definition} \label{def:diamond-tl}
    We define the tail operation on diamonds and single actions as $\textit{tl} : \Diamond(A) \times A \to \mathcal{P}(\Diamond(A))$ s.t.\ given any action $a\in A$ and diamond $\diamond = \langle C_{a}, C_{s} \rangle \in \Diamond(A)$ we have:

    $$\begin{array}{l l}
        \textit{tl}(\langle C_a, C_s \rangle, a ) &= \{\langle C_a[a \mapsto C_a(a) - 1)], C_s \rangle \mid C_a(a) > 0 \}  \\
         & \cup ~ \{ \langle C_a, C_s[as \mapsto C_s(as) - 1, s \mapsto C_s(s) + 1 ] \rangle \mid \\
         & \quad\quad \exists s \in \nomono. C_s(as) > 0\} \\
         & \cup ~ \{ \langle C_a[b \mapsto C_a(b) + k] , C_s[ab^k \mapsto C_s(ab^k) - 1] \rangle \mid \\
         & \quad\quad \exists b \in A, k \in \mathbb{N}_{>0}. C_s(ab^k) > 0\}
    \end{array}$$

    Similarly, we define the tail operation on diamonds and action sequences such that for all diamonds $\diamond \in \Diamond(A)$, actions $a \in A$ and action sequences $s \in A^*$, we have:
    $$\textit{tl}(\diamond, \varepsilon) = \{\diamond\} \text{, and} \quad \textit{tl}(\diamond, as) = \bigcup_{\diamond' \in \textit{tl}(\diamond, a)} \textit{tl}(\diamond', s) \text{.}$$

    Lastly, we define the tail operation on two diamonds, such that given two diamonds $\diamond, \diamond_{\textit{pf}} \in \Diamond(A)$, we have:
    $$\textit{tl}(\diamond, \diamond_{\textit{pf}}) = \left\{ \begin{array}{l l}
        \{ \diamond \}  & \textbf{if}~\diamond_{\textit{pf}} = \emptydiamond\\
        \bigcup_{\tiny
        \begin{array}{l}
             a \in \textit{hd}(\diamond_{\textit{pf}}),\\
             \diamond' \in \textit{tl}(\diamond, a),\\
             \diamond_{\textit{pf}}' \in \textit{tl}(\diamond_{\textit{pf}}, a)
        \end{array}}
        \textit{tl}(\diamond', \diamond_{\textit{pf}}') &\textbf{otherwise} 
    \end{array} \right. $$
\end{definition}

\begin{example} \label{ex:diamond-tl}
    Below are examples of the \textit{tail} mapping respectively using actions, sequences and diamonds as the second argument.
    
    \noindent $\begin{array}{l l  l l}
    \multicolumn{4}{c}{\text{actions:}}\\
    \textit{tl}(ab)^1||(cd)^1, a) &= \{ b^1||(cd)^1\} \quad& \textit{tl}((ab)^1||(cd)^1, e) &= \emptyset \\
    \textit{tl}((ab)^1||(ac)^1, a) & \multicolumn{3}{l}{= \{b^1||(ac)^1,  c^1||(ab)^1\} }\\
    \multicolumn{4}{c}{\text{action sequences:}}\\
    \textit{tl}((ab)^1||(cd)^1, ab) &= \{(cd)^1\} & \textit{tl}((ab)^1||(cd)^1, ac) &= \{b^1||d^1\} \\
    \multicolumn{4}{c}{\text{diamonds:}}\\
    \textit{tl}(a^4, a^2) &= \{a^2\} & \textit{tl}((ab)^1||(cd)^1, a^1||c^1) &= \{ b^1||d^1\} \\
    \textit{tl}(a^1||b^1, a^1||b^1) &= \{ \emptydiamond \} &
    \textit{tl}((abb)^1||(abd)^1, a^1||(ab)^1) &= \{ b^1||(bd)^1, b^2||d^1 \} \\
    \end{array}$
\end{example}

In Definition \ref{def:sequences-and-diamonds} we define the \emph{sequence-of} relation which indicates if a given sequence is an interleaving of the actions and sequences of a given diamond.
The relation does not enforce that all actions and the complete sequences are contained in the interleaving sequence.
For example, we have $acb \sqin ab||cd$, and $c \sqin ab||cd$, but not $ad \sqin ab||cd$.

\begin{definition} \label{def:sequences-and-diamonds}
    We define the sequence-of relation $(\sqin) : A^* \times \Diamond(A)$ such that:
    \begin{itemize}
        \item $\varepsilon \sqin \diamond$ for any diamond $\diamond \in \Diamond(A)$, and
        \item for any action $a \in A$, sequence $s \in A^*$, and diamond $\diamond \in \Diamond(A)$ if $s \sqin \textit{tl}(\diamond,a)$ then $as \sqin \diamond$. 
    \end{itemize}
\end{definition}

In Definition \ref{def:prefix-diamond} we define when a diamond is considered to be the prefix of another diamond.

\begin{definition}\label{def:prefix-diamond}
    We define the prefix relation on diamonds $\sqsubseteq~ \subseteq \Diamond(A) \times \Diamond(A)$, s.t.\ given any two diamonds $\diamond,\diamond' \in \Diamond(A)$, we have $\diamond \sqsubseteq \diamond'$ iff $\textit{tl}(\diamond, \diamond') \neq \emptyset$.
\end{definition}

In Definition \ref{def:diamond-lts}, we inductively define the precise condition for when we consider a diamond pattern to occur within an LTS.
Informally, we say that a state diamond converges in another state iff every \textit{head}-action of the diamond is enabled, and after taking said action, only the remaining diamond is possible.
This is repeated until the diamond is empty and we have arrived in the target state.
After taking the first transition, the requirement on what actions can be enabled is strengthened (from $\xrightarrow{\diamond}$ to $\xLongrightarrow{\diamond}$) to ensure that only actions pertaining to the current diamond are enabled.
As such, multiple diamonds can be enabled in a given state, but we still enforce exclusivity during the execution of a given diamond.
Unless otherwise specified when talking about diamond convergence, the equivalence relation $=_x$ is assumed to be the standard identity equivalence on states $(=)$.
We list some examples of the diamond convergence relations in Example \ref{ex:diamond-convergence}.

\begin{definition} \label{def:diamond-lts}
    Given an LTS $\langle Q, q_0, A, \to \rangle$, an equivalence relation $=_x$ on states, states $\hat{q}, \check{q} \in Q$, and diamond pattern $\diamond \in \Diamond(A)$ we say that $\hat{q}$ \textit{strictly} $\diamond$-\textit{converges up to} $=_x$ \textit{in} $\check{q}$, denoted as $\hat{q} \xLongrightarrow{\diamond} [\check{q}]_{=_x}$, iff:
    \begin{enumerate}
        \item if $\diamond = \emptydiamond$ then $\hat{q} =_x \check{q}$, 
        \item if $\diamond \neq \emptydiamond$ then for any action $a \in \textit{hd}(\diamond)$ and diamond $\diamond' \in \textit{tl}(\diamond, a)$ there exists state $q \in Q$ with $\hat{q} \xrightarrow{a} q$ and $q \xLongrightarrow{\diamond'}[\check{q}]_{=_x}$, and
        \item if $\diamond \neq \emptydiamond$ then for any outgoing transition $\hat{q} \xrightarrow{a} q$ there exists diamond $\diamond' \in \textit{tl}(\diamond, a)$ with $q \xLongrightarrow{\diamond'} [\check{q}]_{=_x}$.
    \end{enumerate}
    \noindent We say that $\hat{q}$ $\diamond$-converges up to $=_x$ in $\check{q}$, denoted $\hat{q} \xrightarrow{\diamond} [\check{q}]_{=_x}$, if at least the conditions $1$ and $2$ hold.\\
    If $=_x$ is the standard identity equivalence on states $(=)$, we use the simpler notations $\hat{q} \xLongrightarrow{\diamond} \check{q}$ and $\hat{q} \xrightarrow{\diamond} \check{q}$
\end{definition}

\begin{example} \label{ex:diamond-convergence}
    In the LTSs below we have $\hat{q} \xrightarrow{a^1||b^1} \check{q}$ and $\hat{q} \xrightarrow{b^1||c^1} \check{q}'$.
    We do not have $\hat{p} \xrightarrow{a^1||b^1} \check{p}$, because in $p$ we do not have $p \xLongrightarrow{a} \check{p}$ due to the transition $p \xrightarrow{a} \check{p}'$.
    If we were to have some equivalence relation $=_x$ with $\check{p}=_x\check{p}'$, then the weaker diamond convergence $\hat{p} \xLongrightarrow{a^1||b^1} [\check{p}]_{=_x}$ would hold.

    \centering
    \begin{minipage}{0.3\textwidth}
    \centering
    \begin{tikzpicture}
        \node (q1) at (0,0) {$\hat{q}$};
        \node (q2) at (-1.5,-.75) {.};
        \node (q3) at (-.5, -.75) {.};
        \node (q4) at (.5, -.75) {.};
        \node (q5) at (1.5, -.75) {.};
        \node (q6) at (-1, -1.5) {$\check{q}$};
        \node (q7) at (1, -1.5) {$\check{q}'$};
        \draw[->] (q1) -- ++(-.5,0) -- node[left] {$a$} (q2);
        \draw[->] (q1) -- node[pos=.2,left] {$b$} (q3);
        \draw[->] (q1) -- node[pos=.2,right] {$b$} (q4);
        \draw[->] (q1) -- ++(.5,0) --node[right] {$c$} (q5);
        \draw[->] (q2) -- node[left] {$b$} (q6);
        \draw[->] (q3) -- node[pos=.2,left] {$a$} (q6);
        \draw[->] (q4) -- node[pos=.2, right] {$c$} (q7);
        \draw[->] (q5) -- node[right] {$b$} (q7);
    \end{tikzpicture}
    \subexamplecaption{}
    \label{fig:diamond-convergence-l}
    \end{minipage}
    \begin{minipage}{0.3\textwidth}
    \centering
    \begin{tikzpicture}
        \node (p1) at (4,0) {$\hat{p}$};
        \node (p2) at (3.5,-.75) {.};
        \node (p3) at (4.5,-.75) {$p$};
        \node (p4) at (4,-1.5) {$\check{p}$};
        \node (p5) at (5,-1.5) {$\check{p}'$};
        \draw[->] (p1) -- node[pos=.2, left] {$a$} (p2);
        \draw[->] (p1) -- node[pos=.2, right] {$b$} (p3);
        \draw[->] (p2) -- node[left] {$b$} (p4);
        \draw[->] (p3) -- node[pos=.2, left] {$a$} (p4);
        \draw[->] (p3) -- node[pos=.2, right] {$a$} (p5);
    \end{tikzpicture}
    \subexamplecaption{}
    \label{fig:diamond-convergence-r}
    \end{minipage}
\end{example}

With diamond convergence now properly defined, we proceed to prove some properties that one might prefer to follow from diamond convergence.
In Theorem \ref{thm:diamond-contains-all-sequences}, we prove that diamond convergences are complete, i.e.\ all interleavings of the diamond can occur.

\begin{theorem}
\label{thm:diamond-contains-all-sequences}
    Given an LTS $\langle Q, q_0, A, \to \rangle$, states $\hat{q}, \check{q} \in Q$ and diamond $\diamond \in \Diamond(A)$ with $\hat{q} \xrightarrow{\diamond} \check{q}$, we have that for all non-empty sequences $s \in A^*$ with $s ~{\sqin} \diamond$, there exist some state $q' \in Q$ and diamond $\diamond_{\textit{tl}} \in \textit{tl}(\diamond, s)$ such that $\hat{q} \xrightarrow{s} q' \xLongrightarrow{\diamond_{\textit{tl}}} \check{q}$.
\end{theorem}
\noindent \textit{Proof.} Let $\hat{q},\check{q}$ and $\diamond \in \Diamond(A)$ with $\hat{q} \xrightarrow{\diamond} \check{q}$.
We prove using strong induction over $|s|$ that for all non-empty sequences $s \in A^*$ if $s \sqin \diamond$ then there is some state $q' \in Q$ and diamond $\diamond_{\textit{tl}}$ such that $\hat{q} \xrightarrow{s} q' \xLongrightarrow{\diamond_{\textit{tl}}} \check{q}$.
As such, let $k \in \mathbb{N}$ s.t.\ for all $s' \sqin \diamond$ with $0 < |s'| < k$ there exists state $q' \in Q$ and diamond $\diamond' \in \textit{tl}(\diamond, s')$ with $\hat{q} \xrightarrow{s'} q' \xLongrightarrow{\diamond'} \check{q}$, and let $s \in A^*$ with $s \sqin \diamond$ and $|s| = k$.
We make a case distinction on $|s|$. \\
Case $|s| = 1$, we have $s = a$ for some action $a \in A$.
Since $a \in \textit{hd}(\diamond)$, there exists diamond $\diamond_{\textit{tl}} \in \textit{tl}(\diamond, a)$.
Since $\hat{q} \xrightarrow{\diamond} \check{q}$, there exists state $q \in Q$ such that $\hat{q} \xrightarrow{a} q \xLongrightarrow{\diamond_\textit{tl}} \check{q}$. \\
Case $|s| > 1$, there exist $a \in A, s' \in A^*$ with $s = s'a$.
Since $|s'| < k$, there exist $q' \in Q, \diamond' \in \Diamond(A)$ with $\hat{q} \xrightarrow{s} q' \xLongrightarrow{\diamond'} \check{q}$.
Since $q' \xLongrightarrow{\diamond'} \check{q}$ and $a \in \textit{hd}(\diamond')$, we have $q'' \in Q, \diamond'' \in \textit{tl}(\diamond, s'a)$ with $\hat{q} \xrightarrow{s'} q' \xrightarrow{a} q'' \xLongrightarrow{\diamond''} \check{q}$. \qed 
\vspace{0.4em}

Within the context of finding and replacing diamonds in a given LTS, it is important that diamonds cannot overlap.
If this were possible, and a particular diamond and its transitions and states were replaced with a single diamond transition, then any other diamond that shares transition with the replaced diamond would no longer be enabled since part of its transitions are now gone, as is illustrated in Example \ref{ex:overlapping-maximal-diamonds}.
Towards ensuring that the above does not happen, in Theorem \ref{thm:no-overlapping-maximal-diamonds} we state that if two strict diamond convergences occur from a given state, then one of the two diamonds is a prefix of the other diamond.
As such, as long as only the biggest diamond is taken, no information is lost when replacing diamonds.

\begin{example} \label{ex:overlapping-maximal-diamonds}
    In the LTS below, the distinct diamonds $\diamond_1$ and $\diamond_2$ share an overlap in states, i.e.\ $q_{\textit{pf}}$ and all other states belonging to the action sequence $s_{\textit{pf}}$.
    If the transitions and states belonging to $\diamond_1$ were replaced with a single diamond transition, the part of the behaviour also belonging to $\diamond_2$, i.e.\ the $s_{\textit{pf}}$ transition, would disappear, and $\hat{q} \xrightarrow{\diamond_2} q_2$ would no longer hold.
    
    \centering
    \begin{tikzpicture}
        \node (qi) at (0,0) {$\hat{q}$};
        \node (qpf) at (0,-1) {$q_{\textit{pf}}$};
        \node (q1) at (-2, -1.5) {$q_1$};
        \node (q2) at (2, -1.5) {$q_2$};

        \draw[->] (qi) -- node[pos=.5, right] {$s_{\textit{pf}}$} (qpf);
        \draw[double, ->] (qpf) -- node[pos=.3, above] {$\diamond_1'$} (q1);
        \draw[double, ->] (qpf) -- node[pos=.3, above] {$\diamond_2'$} (q2);
        \draw[->] (qi) -- node[pos=.2, left] {$\diamond_1$} (q1);
        \draw[->] (qi) -- node[pos=.2, right] {$\diamond_2$} (q2);

        %\node[dashed, draw=gray, label=$\diamond_1$, fit=(qi) (qpf) (q1)] {};
        %\node[dashed, draw=gray, label=$\diamond_2$, fit=(qi) (qpf) (q2)] {};
    \end{tikzpicture}
\end{example}

\begin{theorem}
\label{thm:no-overlapping-maximal-diamonds}
    Given any LTS $\langle Q, q_0, A, \to \rangle$, states $q, q_1, q_2 \in Q$, and diamonds $\diamond_1, \diamond_2 \in \Diamond(A)$, we have that if $q \xLongrightarrow{\diamond_1} q_1$ and $q \xLongrightarrow{\diamond_2} q_2$ then either $\diamond_1 \sqsubseteq \diamond_2$ or $\diamond_2 \sqsubseteq \diamond_1$.
\end{theorem}
\noindent \textit{Proof.} We prove this using strong induction on $|\diamond_1|$. 
Let $k \in \mathbb{N}$ and let us assume, as our induction hypothesis, that for all states $q,q_1, q_2 \in Q$ and diamonds $\diamond_1, \diamond_2 \in \Diamond(A)$ with $q \xLongrightarrow{\diamond_1} q_1$, $q \xLongrightarrow{\diamond_2} q_2$ , and $|\diamond_1| < k$ we have that either $\diamond_1 \sqsubseteq \diamond_2$ or $\diamond_2 \sqsubseteq \diamond_1$.
Now, let us assume states $q,q_1,q_2 \in Q$, and diamonds $\diamond_1, \diamond_2 \in \Diamond(A)$ with $q \xLongrightarrow{\diamond_1} q_1$, $q \xLongrightarrow{\diamond_2} q_2$ , and $|\diamond_1| = k$.
We proceed with case distinction on $|\diamond_1|$.

\textit{Case} $|\diamond_1| = 0$. As such $\diamond_1 = \emptydiamond \sqsubseteq \diamond_2$. 

\textit{Case} $|\diamond_1| > 0$. Thus there exist some action $a \in \textit{hd}(\diamond_1)$ and diamond $\diamond_1' \in \textit{tl}(\diamond_1, a)$.
It follows that there exists some state $q' \in Q$ with $q \xrightarrow{a} q' \xLongrightarrow{\diamond_1'} q_1$.
Since $q \xLongrightarrow{\diamond_2} q_2$, we have that $a \in \textit{hd}(\diamond_2)$ and there exists diamond $\diamond_2' \in \textit{tl}(\diamond_2, a)$ with $q' \xLongrightarrow{\diamond_2'} q_2$.
Since $|\diamond_1'| < k$, we have, as per our induction hypothesis, that either $\diamond_1' \sqsubseteq \diamond_2'$ or $\diamond_2' \sqsubseteq \diamond_1'$.

If $\diamond_1' \sqsubseteq \diamond_2'$, then $\textit{tl}(\diamond_2', \diamond_1') \neq \emptyset$.
Consequently, since $a \in \textit{hd}(\diamond_1)$, $\diamond_2' \in \textit{tl}(\diamond_2, a)$, and $\diamond_1' \in \textit{tl}(\diamond_1, a)$, we have that $\textit{tl}(\diamond_2, \diamond_1) \neq \emptyset$, and thus $\diamond_1 \sqsubseteq \diamond_2$.
In the same manner, we have that if $\diamond_2' \sqsubseteq \diamond_1'$ then $\diamond_2 \sqsubseteq \diamond_1'$. \qed
\vspace{0.4em}

\section{Diamond Preservation under Equivalence Relations} \label{sec:diamond-preservation}
As we have remarked before, reduction modulo some equivalence relation is a useful tool when it comes to reducing the complexity of state spaces.
In Definition \ref{def:diamond-equiv}, we define the notion of \emph{diamond equivalence}, i.e.\ two states are diamond equivalent iff the same diamond convergences are possible in both states.
A desirable property would be for diamond equivalence to be preserved by a given equivalence reduction, e.g.\ reduction modulo some equivalence relation neither adds nor removes diamond convergences.
What we find, is that diamond-equivalence and strong bisimulation are in fact equivalent.
In Theorem \ref{thm:diamond-equiv-implies-bisim} we show that a diamond equivalence relation is a bisimulation, and conversely, in Theorem \ref{thm:bisim-implies-diamond-equiv} we show that a bisimulation relation is a diamond equivalence relation.

\begin{definition} \label{def:diamond-equiv}
Given an LTS $\langle Q, q_0, A, \to \rangle$, and some equivalence relation $R \subseteq Q \times Q$, we say that $R$ is a diamond equivalence relation iff:
    For all states $\hat{q}, \check{q}, \hat{q}' \in Q$ and diamonds $\diamond \in \Diamond(A)$ if $\hat{q} {R} \hat{q}'$ and $\hat{q} \xrightarrow{\diamond} \check{q}$ then $\hat{q}' \xrightarrow{\diamond} [\check{q}]_R$.
\end{definition}

\begin{corollary}
    Given any diamond equivalence relation $=_\diamond$, LTSs $l = \langle Q, q_0, A,$\\$ \to \rangle$, and $l' = \langle Q', q_0', A \to' \rangle$ s.t.\ $q_0 =_\diamond q_0'$ and $l'$ is minimal modulo $=_{\diamond}$. 
    Then for all states $\hat{q},\check{q} \in Q, \hat{q}',\check{q}' \in Q'$ with $\hat{q} =_\diamond \hat{q}'$ and $\check{q} =_\diamond \check{q}'$ we have if $\hat{q}$ $\diamond$-converges to $\check{q}$ for some diamond $\diamond \in \Diamond(A)$, then $\hat{q}'$ $\diamond$-converges to $\check{q}'$.
\end{corollary}

\begin{theorem} \label{thm:diamond-equiv-implies-bisim}
    Any diamond equivalence relation is a bisimulation.
\end{theorem}
\noindent \textit{Proof.} Let $\langle Q,q_0,A,\to\rangle$ be some LTS, $\hat{q}, \hat{q}', \check{q}' \in Q$ be states, $=_\diamond$ some diamond equivalence relation, and $a \in A$ be some action such that $\hat{q} =_\diamond \hat{q}'$ and $\hat{q} \xrightarrow{a} \check{q}$.
We now prove that there is some $\check{q}' \in Q$ such that $\hat{q}' \xrightarrow{a} \check{q}'$ and $\check{q} =_\diamond \check{q}'$.
We note that $a^1 \in \Diamond(A)$ and since $\hat{q} \xrightarrow{a} \check{q}$, we have that $\hat{q}$ $a^1$-converges in $\check{q}$.
As per the $=_\diamond$ equivalence relation, there exists state $\check{q}' \in Q$ such that $\hat{q}'$ $a^1$-converges up to $=_\diamond$ in $\check{q}'$.
Or in other words, there exists a state $\check{q}' \in Q$ with $\hat{q} \xrightarrow{a} \check{q}$ and $\check{q} =_\diamond \check{q}'$.\qed

\begin{theorem} \label{thm:bisim-implies-diamond-equiv}
    Any strong bisimulation relation is a diamond-equivalence relation.
\end{theorem}
\noindent \textit{Proof.} Let us assume some LTS $l = \langle Q, q_0, A, \to \rangle$, and strong bisimulation relation $ \bisim$.
We prove that for all states $\hat{q}, \hat{q}', \check{q} \in Q$ and $\diamond \in \Diamond(A)$ with $\hat{q}\bisim \hat{q}'$ we have that if $\hat{q} \xrightarrow{\diamond} \check{q}$ then $\hat{q}' \xrightarrow{\diamond} [\check{q}]_{ \bisim}$, and if $\hat{q} \xLongrightarrow{\diamond} \check{q}$ then $\hat{q}' \xLongrightarrow{\diamond} [\check{q}]_{ \bisim}$ using strong induction on $|\diamond|$.
We note that diamond equivalence follows from the first implication.

Let us thus assume some $k \in \mathbb{N}$ s.t.\ the above holds for all states $\hat{q},\hat{q}',\check{q} \in Q$ and diamond $\diamond \in \Diamond(A)$ with $|\diamond| < k $.
We now prove that given states $\hat{q},\hat{q}',\check{q} \in Q$ with $\hat{q}  \bisim \hat{q}'$ and diamond $\diamond \in \Diamond(A)$ with $|\diamond| = k$ and $\hat{q} \xrightarrow{\diamond} \check{q}$, we have that if $\hat{q} \xrightarrow{\diamond} \check{q}$ then $\hat{q}' \xrightarrow{\diamond} [\check{q}]_{ \bisim}$, and if $\hat{q} \xLongrightarrow{\diamond} \check{q}$ then $\hat{q}' \xLongrightarrow{\diamond} [\check{q}]_{ \bisim}$.

\textit{Case} $k = 0$, i.e.\  $\diamond = \emptydiamond$.
Thus we have that $\hat{q} = \check{q}$, and consequently $\check{q}  \bisim \hat{q}'$.
Since $\diamond = \emptydiamond$, we have that $\hat{q}' \xrightarrow{\diamond} \hat{q}'$ and $\hat{q}' \xLongrightarrow{\diamond} \hat{q}'$, and thus $\hat{q}' \xrightarrow{\diamond} [\check{q}]_{ \bisim}$ and $\hat{q}' \xLongrightarrow{\diamond} [\check{q}]_{ \bisim}$.

\textit{Case} $k > 0$.
Let us assume that $\hat{q} \xrightarrow{\diamond} \check{q}$, and let $a \in \textit{hd}(\diamond)$ and $\diamond_{\textit{tl}} \in \textit{tl}(\diamond, a)$.
It follows that there exists a state $q \in Q$ with $\hat{q} \xrightarrow{a} q \xLongrightarrow{\diamond_\textit{tl}} \check{q}$.
Since $\hat{q}  \bisim \hat{q}'$, there exists a state $q' \in Q$ with $\hat{q}' \xrightarrow{a} q'$ and $q  \bisim q'$.
Since $|\diamond_\textit{tl}| < k$, we have that $q' \xLongrightarrow{\diamond_\textit{tl}} [\check{q}]_{\bisim}$, and consequently $\hat{q}' \xrightarrow{\diamond} [\check{q}]_{\bisim}$

Now let us make the stronger assumption that $\hat{q} \xLongrightarrow{\diamond} \check{q}$, and let  $q \in Q$ and $a \in A$ s.t.\ $\hat{q} \xrightarrow{a} q$.
Since $\hat{q}  \bisim \hat{q}'$, there exists a state $q' \in Q$ with $\hat{q}' \xrightarrow{a} q'$ and $q  \bisim q'$.
Since $\hat{q} \xLongrightarrow{\diamond} \check{q}$, we have $\diamond_\textit{tl} \in \textit{tl}(\diamond, a)$ s.t.\ $q \xLongrightarrow{\diamond_\textit{tl}} \check{q}$.
Since $|\diamond_\textit{tl}| < k$, we have that $q' \xLongrightarrow{\diamond_\textit{tl}} [\check{q}]_{ \bisim}$, and consequently $\hat{q}' \xLongrightarrow{\diamond} [\check{q}]_{\bisim}$

It thus follows that any strong bisimulation relation is also a diamond-equivalence relation.
\qed

Unlike strong bisimilarity, the weaker equivalence relation of trace-equivalence is not diamond-preserving, as showcased by Example \ref{ex:weak-trace-counter}.
Since trace equivalence and bisimulation are equivalent for deterministic systems, it follows that trace equivalence does preserve diamonds given that the LTS is deterministic.

\begin{example} \label{ex:weak-trace-counter}
    Consider the LTSs below, we have that states $p$ and $q$ are trace equivalent, as $\textit{traces}(p) = \textit{traces}(q) = \{ba_1a_2, a_1ba_2, a_1a_2b \}$, however the diamond $b^1 || (a_1a_2)^1$ is not preserved, since the actions $a_1$ and $a_2$ are not enabled in the state $q'$. 
    
    \centering
    \begin{minipage}{0.4\textwidth}
    \centering
    \begin{tikzpicture}
    \node (q00) at (0,0) {$p$};
    \node (q01) at (.75,-.75) {$.$};
    \node (q02) at (1.5,-1.5) {$.$};
    \node (q10) at (-.75,-.75) {$.$};
    \node (q11) at (0,-1.5) {$.$};
    \node (q12) at (.75,-2.25) {$.$};

    \draw[->] (q00) -- node[pos=.5,  above] {$a_1$} (q01);
    \draw[->] (q01) -- node[pos=.5,  above] {$a_2$} (q02);
    \draw[->] (q00) -- node[pos=.5,  above] {$b$} (q10);
    \draw[->] (q01) -- node[pos=.5,  above] {$b$} (q11);
    \draw[->] (q02) -- node[pos=.5,  above] {$b$} (q12);
    \draw[->] (q10) -- node[pos=.5,  above] {$a_1$} (q11);
    \draw[->] (q11) -- node[pos=.5,  above] {$a_2$} (q12);
    \end{tikzpicture}
    \subexamplecaption{}
    \label{fig:weak-trace-counter-l}
    \end{minipage}
    \begin{minipage}{0.3\textwidth}
    \centering
    \begin{tikzpicture}
    \node(qe) at (0,0) {$q$};

    \node(qb) at (-.75,-.75) {$.$};
    \node(qba1) at (-.75,-1.5) {$.$};
    \node(qa1) at (0, -.75) {$q'$};
    \node(qa1b) at (0, -1.5) {$.$};
    \node(qa12) at (.75,-.75) {$.$};
    \node(qa2) at (.75,-1.5) {$.$};

    \node(qf) at (0,-2.25) {$q'$};

    \draw[->] (qe) -- ++(-.75,-.2) -- node[pos=.5, right] {$b$} (qb);
    \draw[->] (qe) -- node[right] {$a_1$} (qa1);
    \draw[->] (qe) -- ++(.75,-.2) -- node[pos=.5, right] {$a_1$} (qa12);

    \draw[->] (qb) -- node[pos=.5, right] {$a_1$} (qba1);
    \draw[->] (qa1) -- node[pos=.2, right] {$b$} (qa1b);
    \draw[->] (qa12) -- node[pos=.5, right] {$a_2$} (qa1a2);

    \draw[->] (qba1) -- node[pos=.5, right] {$a_2$} ++(0,-.5) -- (qf);
    \draw[->] (qa1b) -- node[pos=.5, right] {$a_2$} (qf);
    \draw[->] (qa1a2) -- node[pos=.5, right] {$b$} ++(0,-.5) -- (qf);
    \end{tikzpicture}
    \subexamplecaption{}
    \label{fig:weak-trace-counter-r}
    \end{minipage}
\end{example}

\color{black}
\section{Finding diamonds} \label{sec:finding-diamonds}
We now outline our algorithm, shown in Algorithm \ref{alg:diamond-global}, for finding all diamonds in a given LTS.
Our algorithm navigates backwards across incoming transitions from a particular origin state $\check{q}$ to find all diamonds converging in $\check{q}$.
The \textit{todo} list of states contains states that have a strict diamond-convergence in $\check{q}$.
States are added to this list in a breadth-first manner, and removed in FIFO order.
Starting with only diamonds consisting of a single action, by repeatedly executing the Step algorithm, shown in Algorithm \ref{alg:diamond-step}, we check to see if a larger diamond, consisting of one additional action, converging in $\check{q}$ exists.
When a larger strict diamond convergence is found, the source state is added to the \textit{todo} list.
When the list is empty, the algorithm terminates.

During execution, both algorithms update two mappings.
The convergence mapping $\Diamond$, keeps track of any found diamond convergence, i.e.\ the diamond and resulting state.
The strictness mapping $x$ is used to keep track of whether or not the outgoing diamond convergence is a strict diamond convergence.
Since multiple strict diamond convergences from a single state cannot occur unless one diamond is a prefix of the other, as a result of Theorem \ref{thm:no-overlapping-maximal-diamonds}, it is sufficient to keep track of this per state, instead of per diamond-convergence.
When the algorithm terminates, the mappings $\Diamond$ and $x$ will have been fully populated.
When used on every state in the LTS, the $\Diamond$ mapping can subsequently be used to construct an LTS containing precisely all diamond convergences.

\begin{algorithm}[H]
\caption{This algorithm finds all non-empty diamond converging in the state $\check{q}$.}
\label{alg:diamond-global}
\KwData{An LTS $\langle Q, q_0, A, \to\rangle$, and a state $\check{q} \in Q$.}

$\textit{todo} \gets []$\;
\For{$\langle q_\textit{src}, a, \check{q} \rangle \in \to$}{
    $\textbf{add}~\langle a, \check{q}\rangle~\textbf{to}~\Diamond(q_\textit{src})$\;
    \If{$|\{\langle a', q' \rangle  \mid q_{\textit{src}} \xrightarrow{a'} q' \}| = 1$}{
        $x(q_\textit{src}) \gets \true$\;
        $\textbf{add}~q_\textit{src}~\textbf{to}~\textit{todo}$\;
    }
}
\For{$q \in \textit{todo}$ }{
    $\textbf{pick}~\langle \diamond_\textit{tl}, q' \rangle \in \Diamond(q)$ \emph{\textbf{with}} $q' = \check{q}$\;
    \For{$\langle \hat{q}, a, q\rangle \in \to$ \emph{\textbf{with}} $\hat{q} \notin \textit{todo}$}{
        \If{$\textit{Step}(\hat{q}, a, q, \diamond_\textit{tl}, \check{q}) = \textit{strict}, \diamond_H$}{
            $\textbf{add}~\hat{q}~\textbf{to}~\textit{todo}$\;
        }
    }
}
\end{algorithm}

We now discuss Algorithm \ref{alg:diamond-global} in more detail.
First, for each incoming transition $q_\textit{src} \xrightarrow{a} \check{q}$, we test to see if this transition is the only outgoing state of the source state.
If so, we have that $q_\textit{src} \xLongrightarrow{a^1} \check{q}$ and $q_\textit{src}$ is added to $\textit{todo}$, so that we can look for any larger diamond converging in $\check{q}$.
Second, the \textit{Step} algorithm, outlined in Algorithm \ref{alg:diamond-step}, is repeatedly executed to answer the following question: given states $\hat{q}, q, \check{q} \in Q$, action $a \in A$, and non-empty diamond $\diamond \in \Diamond(A)$ with $\hat{q} \xrightarrow{a} q \xLongrightarrow{\diamond} \check{q}$, is there some diamond $\diamond_H \in \Diamond(A)$ with $\diamond \in \textit{tl}(\diamond_H,a)$ and $\hat{q}
\xrightarrow{\diamond_H} \check{q}$, and if so is it a strict diamond-convergence.
This is also formalised as Theorem \ref{thm:step-invariant}.
If a strict diamond convergence $\hat{q} \xLongrightarrow{\diamond_H} \check{q}$ is found through the Step algorithm, the source state is added to the \textit{todo} list, so that we can look for an even larger diamond converging in $\check{q}$.

The \textit{Step} algorithm works as follows:
First, all hypothesis diamonds are constructed in the set $\Diamond_H$, and tested for diamond convergence in lines $3$ through $12$.
Here, for all $a \in \textit{hd}(\diamond_H)$, the tail diamonds $\textit{tl}(\diamond_H, a)$, are added to \textit{todo}.
Then all matching outgoing transitions are checked to test if the resulting state has a diamond convergence to $\check{q}$.
If \textit{todo} is not empty afterwards, then there is some action $a \in \textit{hd}(\diamond_H)$ and tail diamond $\textit{diamond}_\textit{tl} \in \textit{tl}(\diamond_H, a)$ for which there is no state $q' \in Q$ with $\hat{q} \xrightarrow{a} q' \xLongrightarrow{\diamond_{\textit{tl}}}$, and thus $\hat{q} \xLongrightarrow{\diamond_H} \check{q}$ does not hold. 

If \textit{b} remained $\true$, we proceed to lines $14$ through $21$ to test if the convergence is strict or not.
This is done by testing each outgoing transition to see if it has a corresponding tail diamond in the resulting state.
Since the algorithm is executed breadth-first, we always have that $\Diamond(q')$ and $x(q')$ have been populated if it does diamond converge in $\check{q}$ with the remaining diamond.

In Theorem \ref{thm:step-invariant}, we prove that given that the algorithm is executed breadth-first, i.e.\ $\Diamond$ and $x$ are populated with all smaller diamonds, then the \textit{Step} algorithm will always update $\Diamond$ and $x$ accordingly.

\begin{theorem} \label{thm:step-invariant}
    Given any LTS $\langle Q, \_, A, \to \rangle$, states $\hat{q}, q, \check{q} \in Q$, action $a \in A$, and diamond $\diamond \in \Diamond(A)$, s.t.\  $\hat{q} \xrightarrow{a} q \xLongrightarrow{\diamond} \check{q}$, and mappings $\Diamond : Q \to \mathcal{P}(\Diamond(A) \times Q), x : Q \to \mathbb{B}$, and we have that $\Diamond$ and $x$ are populated with precisely all diamonds originating in $q'$ smaller than $\diamond$, i.e.\
    $$\forall_{\diamond' \in \Diamond(A), q' \in Q}. |\diamond'| \leq |\diamond| \Rightarrow (q' \xrightarrow{\diamond'} \check{q} \Leftrightarrow \langle \diamond', \check{q} \rangle \in \Diamond(\hat{q})) \wedge (q' \xLongrightarrow{\diamond'} \check{q} \Leftrightarrow x(q'))\text{, then}$$
    the following predicate is an invariant for calling the $\textit{Step}(\hat{q}, a, q, \diamond, \check{q})$ function shown in Algorithm \ref{alg:diamond-step}:
    for all states $\hat{q}' \in Q$ we have
    $$\begin{array}{l l}
        & \forall \langle \diamond', \check{q}' \rangle \in \Diamond(\hat{q}').~ \hat{q}' \xrightarrow{\diamond} \check{q}'\\
        \wedge & x(\hat{q}') \Leftrightarrow \exists \langle \diamond', \check{q}' \rangle \in \Diamond(\hat{q}').~ \hat{q}' \xLongrightarrow{\diamond'} \check{q}' \text{.}\\
    \end{array} $$
\end{theorem}
\noindent \textit{Proof.} Let us assume that the invariant holds initially.
We note the following predicate is an invariant for the \textbf{for}-loops on lines 4 and 6:
$$\neg b \Rightarrow \exists a\in \textit{hd}(\diamond_H), \diamond_\textit{tl} \in \textit{tl}(\diamond_H, a). \neg \exists q'\in Q. \hat{q} \xrightarrow{a} q' \xLongrightarrow{\diamond_\textit{tl}} \check{q}\\
$$
We now prove that when $\langle \diamond_H, \check{q}\rangle$ is added to $\Diamond(\check{q})$ on line $14$, we have $\hat{q} \xrightarrow{\diamond_H} \check{q}$.
Towards this, let us assume some action $a_i \in \textit{hd}(\diamond_H)$ and diamond $\diamond_i \in \textit{tl}(\diamond, a_i)$.
Since $b$ is still $\true$ and we have iterated over all $a \in \textit{hd}(\diamond_H)$, it follows from the aforementioned loop invariant that $\hat{q} \xrightarrow{\diamond_H} \check{q}$, and thus the first part of the invariant for calling $\textit{Step}(\hat{q}, a, q, \diamond, \check{q})$ holds.

Second, we prove that when $x(\check{q})$ is set to $\true$, we also have $\hat{q} \xLongrightarrow{\diamond_H} \check{q}$.
We note that when the \textbf{for}-loop on line 15 completes successfully, i.e.\ \textit{non-strict} is never returned, we have for any $a' \in A, q' \in Q$ with $\hat{q} \xrightarrow{a'}$ that $a' \in \textit{hd}(\diamond_H)$ and $q' \xLongrightarrow{\diamond_{\textit{tl}}} \check{q}$ with $\diamond_{\textit{tl}} \in \textit{tl}(\diamond_H, a')$, and thus we have $\hat{q} \xLongrightarrow{\diamond_H} \check{q}$. 

Lastly, we prove that if $\textit{non-strict}, \diamond_H$ is returned, and $x(\hat{q})$ is not changed, then we have that $\neg (\hat{q} \xLongrightarrow{\diamond_H} \check{q})$.
Since \textit{non-strict} is returned, there exists $a \in A, q' \in Q$ with $\hat{q} \xrightarrow{a} q'$ and either $a'\notin \diamond_H$, $\neg x(q')$, or $\Diamond(q') \cap \{\langle \diamond_{\textit{tl}}, \check{q}\rangle \mid \diamond_\textit{tl} \in \textit{tl}(\diamond_H, a') \}$, let $a, q'$ be as such.
If $a' \in \textit{hd}(\diamond_H)$, then $\neg (\hat{q} \xLongrightarrow{\diamond_H} \check{q})$ trivially follows.
Let us thus assume that $a' \in \textit{hd}(\diamond_H)$ and let us assume towards contradiction, that $\hat{q} \xLongrightarrow{\diamond_H} \check{q}$.
From this assumption it follows that there exists some diamond $\diamond_\textit{tl} \in \textit{tl}(\diamond_H, a')$ with $q' \xLongrightarrow{\diamond_\textit{tl}} \check{q}$.
Since $|\diamond_\textit{tl}| = |\diamond|$, we have that $\langle \diamond_\textit{tl},\check{q} \rangle \in \Diamond(q')$ and $x(q')$.
This contradicts the \textbf{if}-condition evaluating to $\true$, and thus we have $\neg (\hat{q} \xLongrightarrow{\diamond_H} \check{q})$. \qed

\begin{algorithm}[H]
\caption{The Step algorithm finds the possible (exclusive) diamond in a given state, given that only said diamond is available in the given state.}
\label{alg:diamond-step}
\KwData{An LTS $\langle Q, q_0, A, \to\rangle$, a state $q_i \in Q$, states $\hat{q}, q, \check{q} \in Q$, action $a \in A$, and non-empty diamond $\diamond \in \Diamond(A)_{\setminus \emptydiamond}$, s.t.\ $\hat{q} \xrightarrow{a} q \xLongrightarrow{\diamond} \check{q}$, and mappings $\Diamond : Q \to \mathcal{P}(\Diamond(A) \times Q), x : Q \to \mathbb{B}$}
$\Diamond_H \gets \{\diamond_H \in \Diamond(A) \mid \diamond \in \textit{tl}(\diamond_H, a) \}$\;
\For{$\diamond_H \in \Diamond_H$}{
    $\textit{b} \gets \true$\;
    \For{$a \in \textit{hd}(\diamond_H)$}{
        $\textit{todo} \gets \textit{tl}(\diamond_H, a)$\;
        \For{$q' \in Q$ \emph{\textbf{where}} $\hat{q} \xrightarrow{a_i} q' \wedge x(q')$}{
            $\textit{todo} \gets \textit{todo} \setminus \{\diamond \in \Diamond(A) \mid \langle \diamond, \check{q} \rangle \in \Diamond(q')\}$
        }
        \If{$\textit{todo} \neq \emptyset$}{
            $b \gets \false$\;
        }
    }
    \If{$\textit{b}$}{
        \textbf{add} $\langle\diamond_H, \check{q}\rangle$ \textbf{to} $\Diamond(\hat{q})$\;
        \For{$\langle \hat{q}, a', q' \rangle \in \to$}{
            \If{$a' \notin \textit{hd}(\diamond_H) \vee \neg x(q') \vee \Diamond(q') \cap \{\langle \diamond_{\textit{tl}}, \check{q}\rangle \mid \diamond_\textit{tl} \in \textit{tl}(\diamond_H, a') \} = \emptyset$}{
                \textbf{return} \textit{non-strict}, $\diamond_H$\;
            }
        }
        $x(\hat{q}) \gets \true$\;
        \textbf{return} \textit{strict}, $\diamond_H$\;
    }
}
\textbf{return} \false\;
\end{algorithm}

\begin{figure}
    \centering
    \includegraphics[width=0.45\linewidth]{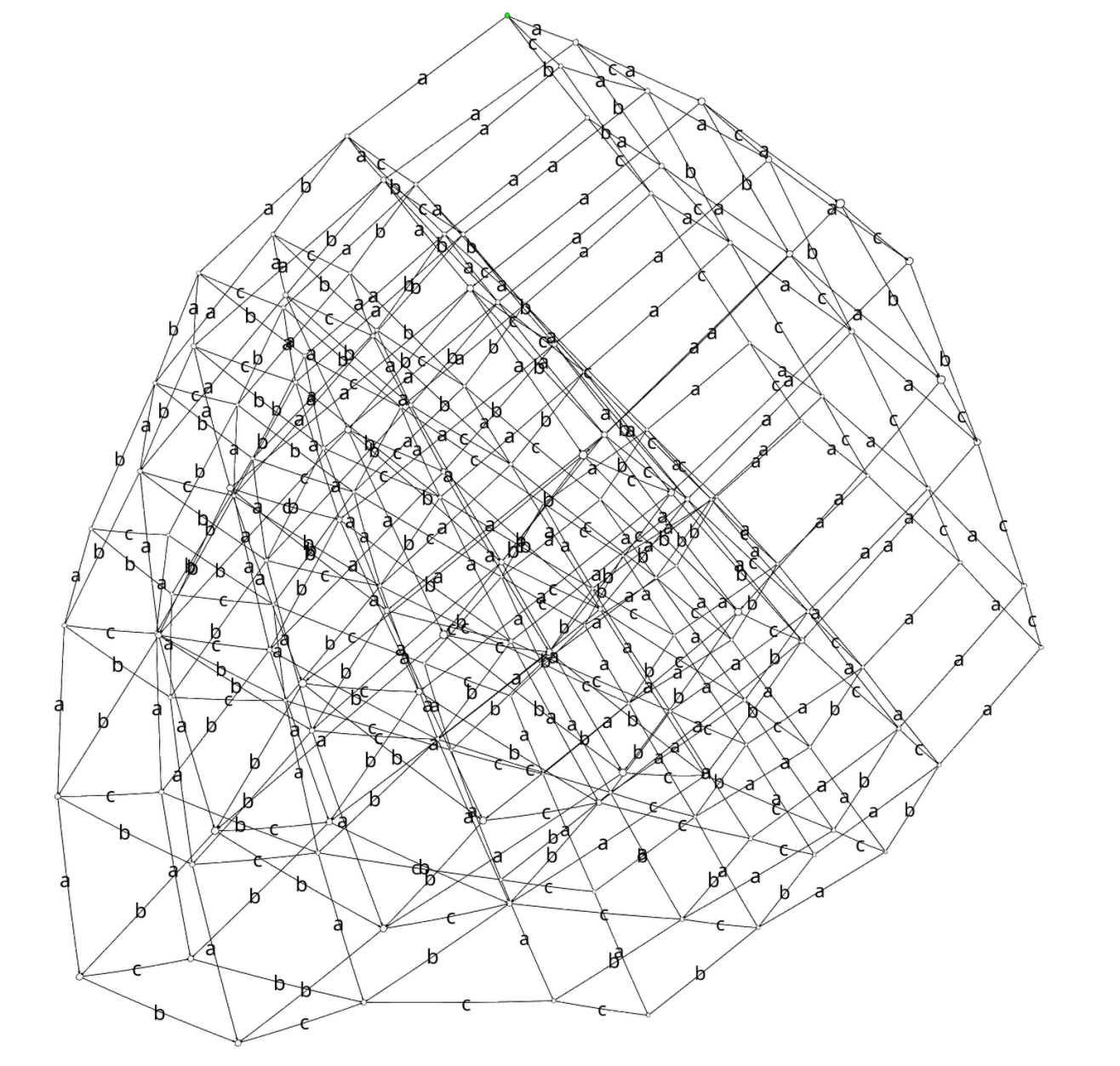}
    \caption{The diamond $a^3 || c^1 || (ab)^2 || (bcc)^1$ before reduction.}
    \label{fig:alg-result}
\end{figure}

A simple proof-of-concept implementation of the algorithm in python is available together with some examples on \cite{jilissen_2025_14980766}.
In Figure \ref{fig:alg-result} we showcase a small toy LTS before applying our algorithm, which finds the given diamond and replaces it with a single transition.
The algorithm was also used on a model of Sliding window protocol, this reduced the number of transitions in the model from 57k transitions to 50k transitions.

Whilst further work towards both space and time complexity of said algorithms is required, we believe the time complexity of the \textit{Step} to be $\mathcal{O}(\lambda^3)$, and the time complexity of the traversal algorithm to be $\mathcal{O}(m*\lambda^3 + n)$, where $m$ is the number of transitions, $n$ the number of states, and $\lambda$ is the largest diamond width in the given LTS.
Since $\lambda$ is generally insignificant compared to $m$, the algorithm has a general runtime of $\mathcal{O}(m + n)$.

\section{Conclusion \& Future Work} \label{sec:conclusion}
We have formally defined the notion of a diamond and when a diamond occurs in a given LTS.
We have proven important properties related to what we desire of the diamond-convergence relation, i.e.\ we have shown that diamonds contain all possible interleavings of their sequences and that a maximal diamond in a given state can safely be replaced with a single diamond transition.
Additionally, we have proven that strong bisimulation is the precise equivalence relation that preserves diamonds for arbitrary LTSs.
Lastly, we have introduced a novel algorithm, with correctness proof, for finding all maximal diamonds in any LTS.

The algorithm and techniques introduced in this paper capture precisely all diamond convergences in any arbitrary LTS.
We stress that capturing precisely all such convergences is a novel part of this work that is not covered by partial order reduction techniques, which present the state-of-the-art techniques for dealing with such patterns.
In particular, not only the presence of a diamond convergence but also the lack thereof can provide additional insight to the user.

We believe the following continuations to be of particular interest for future work:
In model-based techniques, it is convenient to abstract away internal operations using the $\tau$ action.
It would be intuitive to further refine the theory surrounding diamonds in relation to the special $\tau$ action and equivalence relations associated with it.
Other future work exists in further analysis of the time and space complexity of the introduced algorithm, and further optimizations.
In particular, when calculating the diamond convergences for every state, transitions are often considered multiple times.
Reducing repeated work could be a prime candidate for optimization. 

% BEGIN ADDITION

\bibliographystyle{splncs04}
\bibliography{bibliography}

\end{document}